\documentclass[preprint,review,12pt]{elsarticle}

\usepackage{amssymb}
\usepackage{subfigure}
\usepackage{amsmath}

\journal{Optics Communications}

\newcommand{\epl}{Europhys. Lett.\ }

\newcommand{\pra}{Phys. Rev. A\ }
\newcommand{\prl}{Phys. Rev. Lett.\ }
\newcommand{\pr}{Phys. Rev.\ }

\newcommand{\jpa}{J. Phys. A\ }
\newcommand{\jpb}{J. Phys. B\ }
\newcommand{\etal}{{\em et al.}\ }
\newcommand{\e}{\mbox{e}}

\newcommand{\njp}{New J. Phys.\ }
\newcommand{\oc}{Opt. Commun.\ }
\newcommand{\josab}{J. Opt. Soc. Am. B\ }

\newcommand{\UQ}{School of Mathematics and Physics, University of Queensland, Brisbane, 
QLD 4072, Australia.}

\begin{document}

\begin{frontmatter}

\title{Quantum behaviour of pumped and damped triangular Bose Hubbard systems}

\author[cvc]{C.V.~Chianca}
\author[mko]{M.K.~Olsen\corref{cor1}}
\ead{mko@physics.uq.edu.au}

\cortext[cor1]{Corresponding author}

\address{\UQ}

\begin{abstract}

We propose and analyse analogs of optical cavities for atoms using three-well Bose-Hubbard models with pumping and losses. We consider triangular configurations.
With one well pumped and one damped, we find that both the mean-field dynamics and the quantum statistics show a quantitative dependence on the choice of damped well. The systems we analyse remain far from equilibrium, preserving good coherence between the wells in the steady-state. We find quadrature squeezing and mode entanglement for some parameter regimes and demonstrate that the trimer with pumping and damping at the same well is the stronger option for producing non-classical states. Due to recent experimental advances, it should be possible to demonstrate the effects we investigate and predict.

\end{abstract}

\begin{keyword}

Bose-Hubbard model, pumped and damped, squeezing, entanglement.

\end{keyword}

\end{frontmatter}

\section{Introduction}
\label{sec:intro}

In this article we analyse pumped and damped three-well triangular Bose-Hubbard models~\cite{BHmodel,Jaksch,BHJoel,Nemoto,Kordas1,Kordas2} in terms of their mean field behaviour and their quantum statistical properties. Our proposal in this work is inspired by recent advances in the techniques of configuring optical potentials~\cite{painting,tylerpaint}, which allow for the fabrication of a variety of geometric configurations. Together with the ability to cause controlled loss from particular lattice sites, utilising either electron beams~\cite{NDC}, or optical methods~\cite{Weitenberg}, this allows for the manufacture and study of innovative lattice configurations. In some ways the pumped and damped systems we investigate here are similar to coupled nonlinear optical cavities~\cite{nlc,Tan}, but there is one very important difference. While not every optical cavity in a cluster need be pumped, it is impossible to construct an optical resonator without losses. As we will demonstrate in what follows, the ability to choose which well is to be damped leads to some very interesting behaviour in both the mean fields and the quantum correlations.  

Driven dissipative dimers with damping at both wells have been analysed by Casteels and Wouters~\cite{WimWouters}, in terms of both entanglement and bistability. Casteels and Ciuti~\cite{WimCiuti} have examined phase transitions in the same system.
Triangular dimers and inline chains with dissipation at one well have also been analysed~\cite{Kordas3,Kordas4}, finding some interesting physical effects.
 Pi\u{z}orn has analysed Bose-Hubbard models with pumping and dissipation~\cite{Pizorn}, using density matrix techniques, which are useful for moderate numbers of atoms and wells. In this work we examine Bose-Hubbard trimers in the triangular configuration, with pumping and dissipation at one well each. A triangular lattice with damping at the middle well, but without pumping, has been analysed by Shchesnovich and Mogilevstev, showing that the mean-field analysis is not accurate~\cite{Shchesnovich}. Since we do not wish to limit ourselves to the smallish number of atoms used in other treatments, we use the truncated Wigner representation~\cite{Graham,Steel}, as in previous works examining different configurations~\cite{BHcav2,BHcav3}. Unlike some of the other methods, the truncated Wigner is easily extensible to higher well and atom numbers, with a system of 11 wells and 2000 atoms having been analysed recently~\cite{Florian}.

\section{Hamiltonian and equations of motion}
\label{sec:model}

In a three-well triangle with pumping and damping at one well each, there are two possible geometric configurations. Labelling the pumped well as number one, we can choose to damp either well one or well two. By symmetry, damping at well three is the same configuration as that with damping at well two.
We begin with the three-well triangular Bose-Hubbard unitary Hamiltonian,
\begin{equation}
{\cal H} = \hbar\chi\sum_{i=1}^{3}\hat{a}_{i}^{\dag\,2}\hat{a}_{i}^{2}-\hbar J \left[\hat{a}_{1}^{\dag}(\hat{a}_{2}+\hat{a}_{3})+\hat{a}_{2}^{\dag}\hat{a}_{3} + h.c. \right],
\label{eq:genHamtri}
\end{equation}
where $\hat{a}_{i}$ is the bosonic annihilation operator for the $i$th well, $\chi$ represents the collisional nonlinearity and $J$ is the tunneling strength.
We will always consider that the pumping  from the larger condensate is at well $1$ and can be represented by the Hamiltonian
\begin{equation}
{\cal H}_{pump} = i\hbar\left(\epsilon\hat{a}_{1}^{\dag}-\epsilon^{\ast}\hat{a}_{1}\right),
\label{eq:pump}
\end{equation}
which is of the form commonly used for the investigation of optical cavities. The basic assumption here is that the first well receives atoms from a coherent condensate, represented by the complex amplitude $\epsilon$, which is much larger than any of the modes in the wells we are investigating, so that it will not become depleted over the time scales of interest.
The damping term for well $i$ acts on the system density matrix as the Lindblad superoperator
\begin{equation}
{\cal L}\rho = \gamma\left(2\hat{a}_{i}\rho\hat{a}_{i}^{\dag}-\hat{a}_{i}^{\dag}\hat{a}_{i}\rho-\rho\hat{a}_{i}^{\dag}\hat{a}_{i}\right),
\label{eq:damp}
\end{equation}
where $\gamma$ is the coupling between the damped well and the atomic bath, which we assume to be unpopulated.
If the lost atoms fall under gravity, we are justified in using the Markov and Born approximations~\cite{JHMarkov}.

Following the usual procedures~\cite{QNoise,DFW}, we may map the master equation for the density operator onto a generalised Fokker-Planck equation in the Wigner representation. This is not a true Fokker-Planck equation because it has third-order derivatives and, although it can be mapped onto stochastic difference equations~\cite{nossoEPL}, these are not easy to integrate. By dropping the third-order terms, usually under the assumption that they are small, we may map the problem onto It\^o stochastic equations~\cite{SMCrispin} in the truncated Wigner representation. As a representative example, the equations for pumping at well $1$ and loss at well $2$ are
\begin{eqnarray}
\frac{d\alpha_{1}}{dt} &=& \epsilon - 2i\chi |\alpha_{1}|^{2}\alpha_{1}+iJ(\alpha_{2}+\alpha_{3}), \nonumber \\
\frac{d\alpha_{2}}{dt} &=& -\gamma\alpha_{2} -2i\chi|\alpha_{2}|^{2}\alpha_{2}+iJ(\alpha_{1}+\alpha_{3}) +\sqrt{\gamma}\eta , \nonumber \\
\frac{d\alpha_{3}}{dt} &=& -2i\chi|\alpha_{3}|^{2}\alpha_{3}+iJ(\alpha_{1}+\alpha_{2}),
\label{eq:cav3tri}
\end{eqnarray}
where $\epsilon$ represents the rate at which atoms enter well $1$, $\gamma$ is the loss rate from the selected well, and $\eta$ is a complex Gaussian noise with the moments $\overline{\eta(t)}=0$ and $\overline{\eta^{\ast}(t)\eta(t')}=\delta(t-t')$. The variables $\alpha_{i}$ correspond to the operators $\hat{a}_{i}$ in the sense that averages of products of the Wigner variables over many stochastic trajectories become equivalent to symmetrically ordered operator expectation values, for example $\overline{|\alpha_{i}|^{2}}=\frac{1}{2}\langle\hat{a}_{i}^{\dag}\hat{a}_{i}+\hat{a}_{i}\hat{a}_{i}^{\dag}\rangle$. The initial states in all wells will be vacuum. We note here that we will use $\epsilon=10$ and $\gamma=J=1$ in all our numerical investigations, while using two values of $\chi$, $10^{-3}$ and $10^{-2}$. The equations for configurations with damping at a different well are found by the simple transfer of the terms involving $\gamma$.

The parameters used here are consistent with known experimental values. Fixing the tunneling rate at $J=1$ sets the scale for all the other parameters. Physically, the pumping rate and the loss rate can be varied by adjusting well geometries and the strength of the method used for outcoupling. $J$ itself can be changed by changes in the well depths and separation. The most difficult parameter to change experimentally would be $\chi$, which is possible using Feshbach resonance techniques~\cite{Feshbach}. Using the published results of Albiez \etal\cite{Albiez} and setting their tunneling equal to one, we find that their $\chi\approx 10^{-4}$ in our units. While this is smaller than what we have used, deeper wells would lower $J$ and give a ratio $\chi/J$ consistent with our two values, or $\chi$ could be changed using Feshbach techniques. By reference to the same article, we can also say that our system is in the regime where the three-mode approximation is valid.

We note here that the truncated Wigner has been chosen as an alternative to the exact positive-P representation~\cite{P+} because this was found to be unstable for the trimer configuration with loss at only one well. For a similar configuration with loss at the two unpumped wells~\cite{BHAx}, the positive-P representation was used, and gave indistinguishable results to those found with the truncated Wigner for the same system. For a dimer system with loss at one well, the truncated Wigner reproduced all the positive-P results for first and second order moments~\cite{NGBHcav2}, while being qualitatively correct for higher moments. Since we are not interested in the investigation of any moments higher than second order in this work, we feel justified in using this method.

\section{Quantities of interest}
\label{sec:interest}      

In this work we are interested in the number of atoms in each mode and the correlation functions which are used to detect squeezing in each mode and entanglement between the modes.
The populations in each well are calculated as $N_{i} = \overline{|\alpha_{i}|^{2}}-\frac{1}{2}$ while the correlations we use to detect quantum statistical properties are constructed from expectation values of moments of the mode operators. 
In order to proceed, we define the atomic quadratures as 
\begin{eqnarray}
\hat{X}_{j}(\theta) &= & \hat{a}_{j}\e^{-i\theta}+\hat{a}_{j}^{\dag}\e^{i\theta},
\label{eqn:Xtheta}
\end{eqnarray}
so that the $\hat{Y}_{j}(\theta)=\hat{X}_{j}(\theta+\pi/2)$.  
Single mode squeezing exists whenever a particular quadrature variance is found to be less than $1$, for any angle. As is well known, one of the effects of a $\chi^{(3)}$ nonlinearity can be to cause maximum squeezing to be found at a non-zero quadrature angle~\cite{nlc}, so that it becomes important to investigate all angles. This is not the case for resonant $\chi^{(2)}$ systems such as second harmonic generation, where the best squeezing is found for $\theta=0$. 

In order to detect bipartite entanglement and inseparability, we will use the Duan-Simon inequality~\cite{Duan,Simon} which states that, for any two separable states,
\begin{equation}
DS_{jk}=V(\hat{X}_{j}+\hat{X}_{k})+V(\hat{Y}_{j}-\hat{Y}_{k}) \geq 4.
\label{eq:DS}
\end{equation}
with any violation of this inequality demonstrating the inseparability of modes $j$ and $k$. Note that the angular dependence has been suppressed here for clarity of notation. 

The next quantum statistical effect is the Einstein-Podolsky-Rosen (EPR) paradox~\cite{Einstein}, also known as EPR-steering~\cite{Erwin,Wiseman}. For a continuous variable pumped and damped system, the usual method for demonstrating the presence of steering is with the Reid criterion~\cite{EPRMDR}. This is based on the fact that the Heisenberg Uncertainty Principle requires that
\begin{equation}
V(\hat{X}_{i})V(\hat{Y}_{i}) \geq 1.
\label{eq:HUPXY}
\end{equation}
Reid defines the inferred quadrature variances of  two modes labelled $i$ and $j$, with an observer of mode $j$ inferring values of mode $i$, as 
\begin{eqnarray}
V_{inf}(\hat{X}_{ij}) = V(\hat{X}_{i})-\frac{\left[V(\hat{X}_{i},\hat{X}_{j})\right]^{2}}{V(\hat{X}_{j})}, \nonumber \\
V_{inf}(\hat{Y}_{ij}) = V(\hat{Y}_{i})-\frac{\left[V(\hat{Y}_{i},\hat{Y}_{j})\right]^{2}}{V(\hat{Y}_{j})},
\label{eq:VXYinf}
\end{eqnarray}
where the $\theta$ dependence is again suppressed, and $V(AB) = \langle AB\rangle-\langle A\rangle\langle B\rangle$. When the product of these two inferred variances is less than one, this means that mode $i$ can be steered by mode $j$.
As shown by Reid, a violation of the inequality signifies a two-mode state which demonstrates the EPR paradox. It can be seen that this criterion is directional, with the ability to swap $i$ and $j$ in Eq.~\ref{eq:VXYinf}. In what follows, we will denote the value of the product of the inferred variances as $EPR_{ij}$ when the quadrature variances of mode $i$ are inferred by measurements at mode $j$. When one of the pair $(EPR_{ij},EPR_{ji})$ is less than one and the other is more than one, we have a phenomenon known as asymmetric Gaussian steering. This property has been predicted in optical~\cite{SFG,Sarah,asymSHG} and atomic systems~\cite{TWTEPR}, and measured in the laboratory~\cite{Handchen}. It is now established that it is a general property, and may also exist for non-Gaussian measurements~\cite{oneway}.
Since states which are steerable are a strict subset of the entangled states, and these are a subset of inseparable states, $EPR_{ij}<1$ also provides proof of entanglement and inseparability. In fact, with a pumped and damped dimer, the Reid criterion proved to detect entanglement that was not detected using the Duan-Simon inequality~\cite{NGBHcav2}.

For three mode inseparability, we use the van-Loock Furusawa inequalities (vLF)~\cite{vLF}. The first of these is 
\begin{equation}
V_{ij} = V(\hat{X}_{i}-\hat{X}_{j})+V(\hat{Y}_{i}+\hat{Y}_{j}+g_{k}\hat{Y}_{k}) \geq 4, 
\label{eq:VLF}
\end{equation}
for which the violation of any two demonstrates tripartite inseparability. The $g_{j}$, which are arbitrary and real, can be optimised~\cite{AxMuzz}, using the variances and covariances, as
\begin{equation}
g_{i} = -\frac{V(\hat{Y}_{i},\hat{Y}_{j})+V(\hat{Y}_{i},\hat{Y}_{k})}{V(\hat{Y}_{i})}.
\label{eq:VLFopt}
\end{equation}
Teh and Reid~\cite{Teh&Reid} have shown that, for mixed states, tripartite entanglement is demonstrated if the sum of the three correlations was less than $8$, with genuine tripartite EPR (Einstein-Podolsky-Rosen)-steering~\cite{Einstein,Erwin,Wiseman} requiring a sum of less than $4$. 
The second set set of vLF inequalities,
\begin{equation}
V_{ijk} = V(\hat{X}_{i}-\frac{\hat{X}_{j}+\hat{X}_{k}}{\sqrt{2}})+V(\hat{Y}_{i}+\frac{\hat{Y}_{j}+\hat{Y}_{k}}{\sqrt{2}}) \geq 4,
\label{eq:VLFijk}
\end{equation}
requires the violation of only one to prove tripartite inseparability. Teh and Reid showed that for mixed states any one of these less than $2$ demonstrates genuine tripartite entanglement, while one of them less than $1$ demonstrates genuine tripartite EPR steering. 

Multipartite EPR-steering has also been investigated, with Wang \etal  showing that the steering of a given quantum mode is allowed when not less than half of the total number of modes take part in the steering group~\cite{halfplus}. In a tripartite system, this means that measurements on two of the modes are needed to steer the third. In order to quantify this, we will use the correlation functions developed by Olsen, Bradley, and Reid~\cite{OBR}. With tripartite inferred variances as
\begin{eqnarray}
V_{inf}^{(t)}(\hat{X}_{i}) &=& V(\hat{X}_{i})-\frac{\left[V(\hat{X}_{i},\hat{X}_{j}\pm\hat{X}_{k})\right]^{2}}{V(\hat{X}_{j}\pm\hat{X}_{k})}, \nonumber \\
V_{inf}^{(t)}(\hat{Y}_{i}) &=& V(\hat{Y}_{i})-\frac{\left[V(\hat{Y}_{i},\hat{Y}_{j}\pm\hat{Y}_{k})\right]^{2}}{V(\hat{Y}_{j}\pm\hat{Y}_{k})}, 
\label{eq:V3inf}
\end{eqnarray}
we define
\begin{equation}
OBR_{ijk} = V_{inf}^{(t)}(\hat{X}_{i})V_{inf}^{(t)}(\hat{Y}_{i}),
\label{eq:OBR}
\end{equation}
so that a value of less than one means that mode $i$ can be steered by the combined forces of modes $j$ and $k$.
On a final note, we mention that all the quantities needed for the correlations above can in principle be measured, either by density (number) measurements or via atomic homodyning~\cite{andyhomo}.

\section{Loss at the second well}
\label{sec:gamma2}

\begin{figure}[tbh]
\centering
\includegraphics[width=0.7\columnwidth]{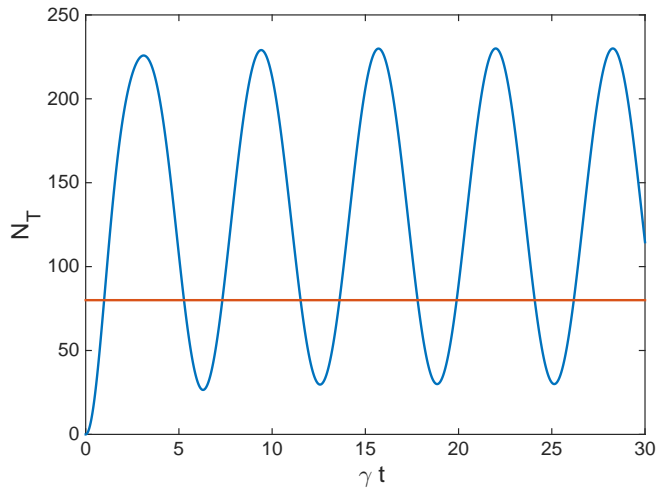}
\caption{(colour online) The classical solutions for the total number of atoms in the system with losses at the second well, compared to the analytical steady-state number, for $J=\gamma=1$, $\epsilon=10$, and $\chi=0$. The numerical solutions do not show a steady-state for wells one and three, although the population in well two does damp to the analytical value. All quantities shown in this and subsequent plots are dimensionless and $\gamma t$ is a dimensionless time.}
\label{fig:NTki0g2}
\end{figure}

With $\chi=0$ we can apparently find analytic steady state classical solutions for the number in each well,
\begin{eqnarray}
N_{1} = N_{2} &=& \frac{|\epsilon|^{2}}{\gamma^{2}+4J^{2}}, \nonumber \\
N_{3} &=& \frac{|\epsilon|^{2}(\gamma^{2}+J^{2})}{J^{2}(\gamma^{2}+4J^{2})},
\label{eq:Ng2class}
\end{eqnarray}
which indicate that the total number of atoms in the system should reach a steady-state. However, when we integrate the equations numerically, we find the result shown in Fig.~\ref{fig:NTki0g2}, where the total number continues to oscillate. Although not shown here, the number in the second well does settle to the steady-state solution, whereas the outer two wells exhibit undamped oscillations at least as far as $\gamma t=200$. This behaviour is reminiscent of self-pulsing in second and third harmonic generation~\cite{SHGpulse,KVKthird,THG} and was also found in the open trimer~\cite{BHcav3}. It prevents us from finding accurate analytical steady state solutions. Numerically we find that the classical oscillations damp out as $\chi$ is increased, but do not disappear completely for the lower $\chi$ value we consider here.

\begin{figure}[tbhp]
\centering
\begin{minipage}{.5\textwidth}
  \centering
  \includegraphics[width=.95\linewidth,height=0.65\linewidth]{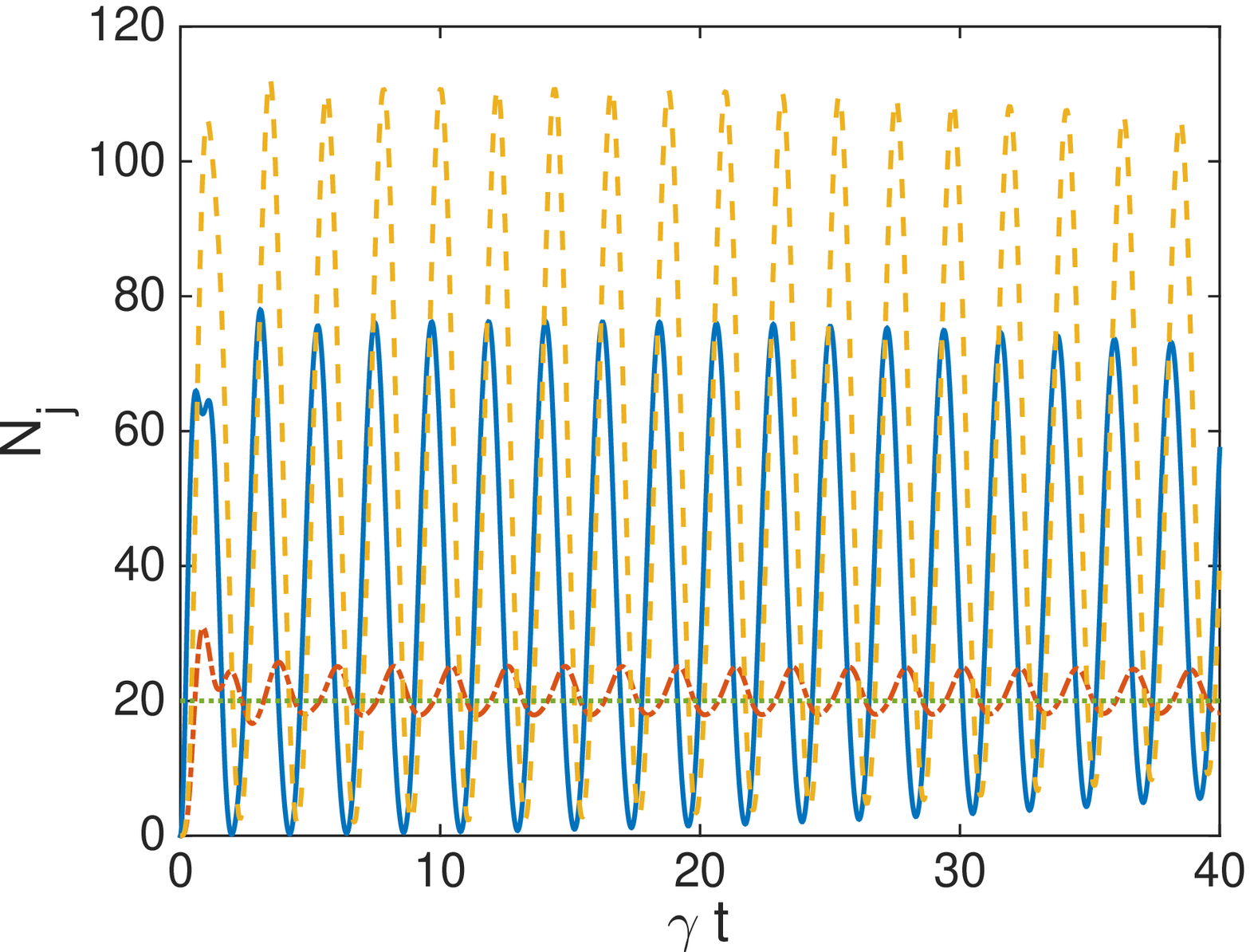}
\end{minipage}%
\begin{minipage}{.5\textwidth}
  \centering
  \includegraphics[width=.95\linewidth,height=0.65\linewidth]{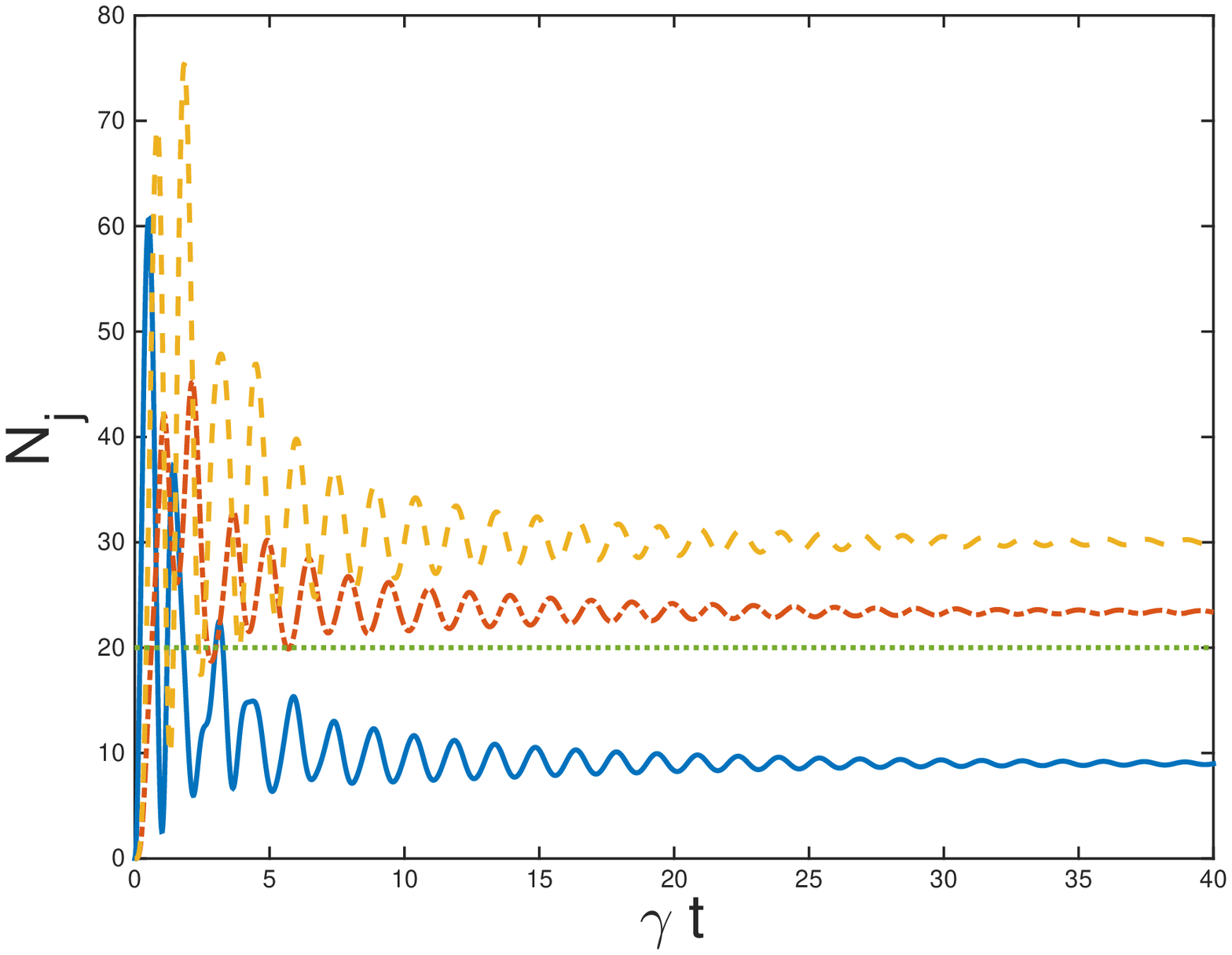}
 \end{minipage}
 \caption{(a) (Color online) The populations in each well of the triangular configuration as a function of time for $\chi=10^{-3}$ and damping at well 2. The solid line is $N_{1}$, the dash-dotted line is $N_{2}$, and the dashed line is $N_{3}$. The dotted line represents the non-interacting steady-state analytical solutions, all of which are equal for this configuration.. \newline
 (b) The populations in each well of the triangular configuration as a function of time for $\chi=10^{-2}$ and damping at well 3. The line styles are the same as in (a).}
 \label{fig:popsg2tri}
\end{figure}

When we integrate the truncated Wigner equations, we find that the oscillations persist over time for $\chi=10^{-3}$, but eventually damp out for $\chi=10^{-2}$, as shown in Fig.~\ref{fig:popsg2tri}. A similar damping is seen in closed trimers~\cite{Chiancathermal} and can be explained by the fact that different number state components of the state vector will have different phase rates due to the Kerr-type nonlinearity. This effect is well known in both the Kerr oscillator and in closed Bose-Hubbard dimers, in which cases the oscillations will revive after some time. Whether a revival will occur at any meaningful time for this pumped and damped system is an open question. We note here that all of the figures to follow are the result of averages of at least $10^{5}$ stochastic trajectories of the truncated Wigner representation, and that samping errors were of the order of the plotted linewidths.

The lack of a steady state for the lower value of the collisional nonlinearity prevents us from finding steady state values for the bipartite correlations which signify squeezing and entanglement. In the case of $\chi=10^{-2}$, we give values found at $\gamma t = 100$, by which time the oscillations had disappeared. As with other $\chi^{(3)}$ systems, the optimum values of these correlations are not found at zero quadrature angle. For this reason, the following tables also give the angle of the optimal correlations. 

\begin{center}
\begin{tabular}{ | c || c | c | c | c | c | c |}
\hline
 \multicolumn{7}{| c |}{Bipartite correlations, loss at 2} \\
 \hline
   &  $V(\hat{X}_{1})$ & $V(\hat{X}_{2})$ & $V(\hat{X}_{3})$ & $DS_{12}$ & $DS_{13}$ & $DS_{23}$  \\ 
 \hline
\hline
 $\chi = 10^{-2}$ & 1.00@178$^o$  & 0.83@131$^o$   & 1.29@75$^o$   & 5.83@164$^o$  & 13.55@53$^o$  & 7.60@86$^o$ \\
 \hline
 \hline
\end{tabular}
\end{center}

We find that wells two and three exhibit quadrature squeezing and that the Duan-Simon criteria do not detect any mode entanglement, with all three results being well above four. However, as shown in the next table, all bipartitions exhibit EPR-steering between the modes, with this being asymmetric for all quadrature angles in the case of wells $2$ and $3$. In the other two bipartitions, the EPR-steering is asymmetric in terms of quadrature angle, with there being no angles where the two modes can steer each other. We note here that the Reid inequalities have again proven to be a more sensitive measure of entanglement than those developed by Duan and Simon. This is possible because the Duan-Simon inequalities are not both necessary and sufficient for mixed state non-Gaussian systems.

\begin{center}
\begin{tabular}{ | c || c | c | c | c | c | c |}
\hline
 \multicolumn{7}{| c |}{Bipartite EPR-steering, loss at 2} \\
 \hline
   &  $EPR_{12}$ & $EPR_{21}$ & $EPR_{23}$ & $EPR_{32}$ & $EPR_{13}$ & $EPR_{31}$  \\ 
 \hline
\hline
 $\chi = 10^{-2}$ & 0.69@175$^o$   & 0.69@131$^o$   & 0.69@133$^o$  & 1.08@80$^o$ & 0.72@2$^o$  & 0.6@71$^o$ \\
 \hline
 \hline
\end{tabular}
\end{center}

The presence of tripartite inseparability and entanglement is not detected by the vLF correlations, all of these satisfying the inequalities. However, as with the bipartite properties, EPR-steering is detected by the products of the inferred variances. This necessarily means that full tripartite entanglement is present in this system. Considering the various categories of tripartite states described by Adesso~\cite{Adesso,OCbitri}, we find that this system produces states of the W type, as both tripartite and bipartite entanglement are found.

\begin{center}
\begin{tabular}{ | c || c | c | c | c | c | c |}
\hline
 \multicolumn{7}{| c |}{Tripartite entanglement, loss at 2} \\
 \hline
   &  $V_{12}$ & $V_{13}$ & $V_{23}$ & $V_{123}$ & $V_{231}$ & $V_{312}$   \\ 
 \hline
 \hline
 $\chi = 10^{-2}$ & 5.72@165$^o$ & 13.04@47$^o$   & 7.60@85$^o$  & 14.22@102$^o$  & 9.35@73$^o$ & 13.90@73$^o$ \\
 \hline
 \hline
\end{tabular}
\end{center}

\begin{center}
\begin{tabular}{ | c || c | c | c |}
\hline
 \multicolumn{4}{| c |}{Tripartite EPR-steering, loss at 2} \\
 \hline
   &   $OBR_{123}$ & $OBR_{231}$ & $OBR_{312}$  \\ 
 \hline
 \hline
 $\chi = 10^{-2}$ & 0.65@180$^o$  & 0.68@131$^o$ & 0.60@76$^o$\\
 \hline
 \hline
\end{tabular}
\end{center}

\section{Losses at the pumped well}
\label{sec:gamma1}

\begin{figure}[tbhp]
\centering
\begin{minipage}{.5\textwidth}
  \centering
  \includegraphics[width=.95\linewidth,height=0.65\linewidth]{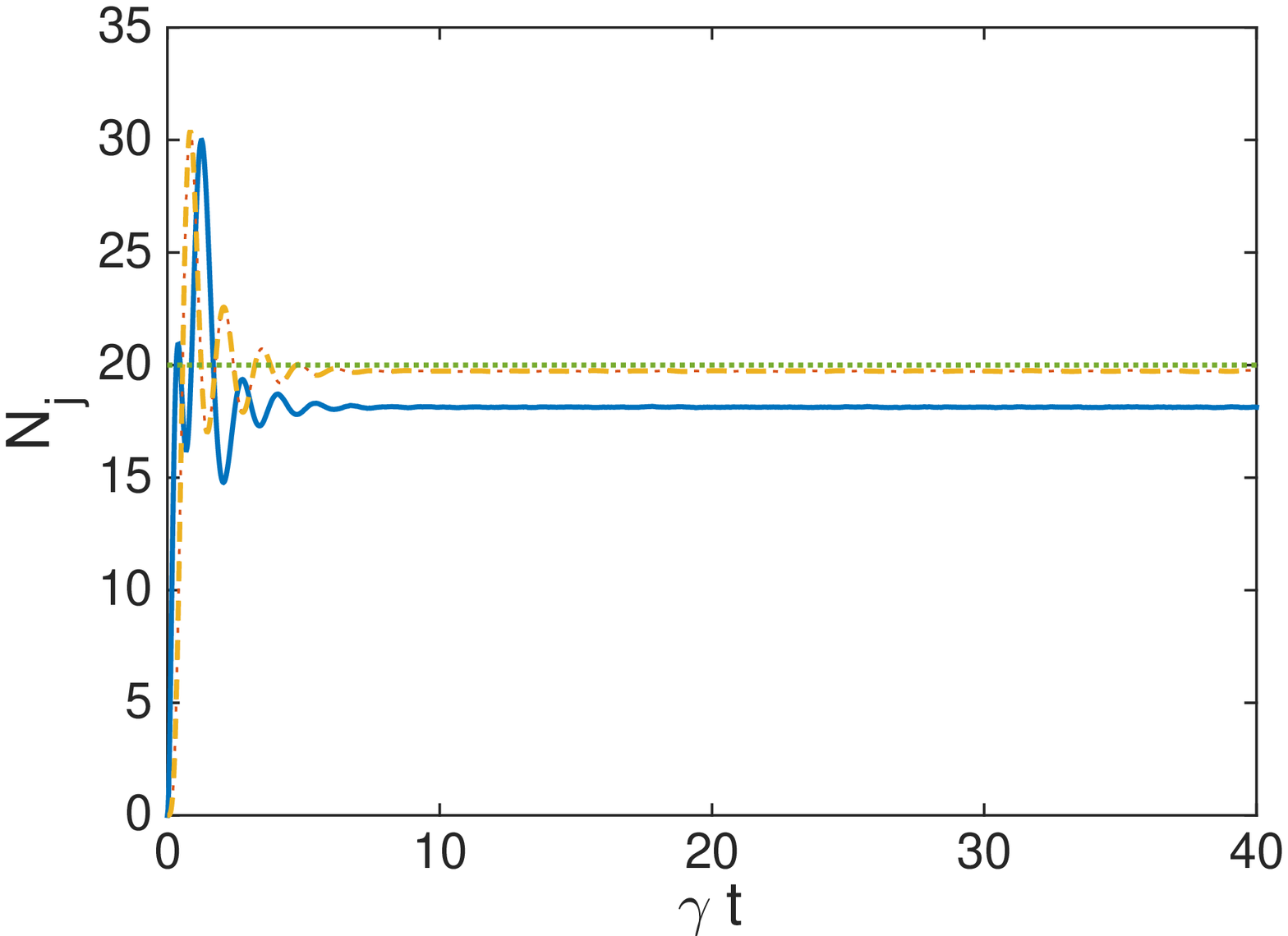}
\end{minipage}%
\begin{minipage}{.5\textwidth}
  \centering
  \includegraphics[width=.95\linewidth,height=0.65\linewidth]{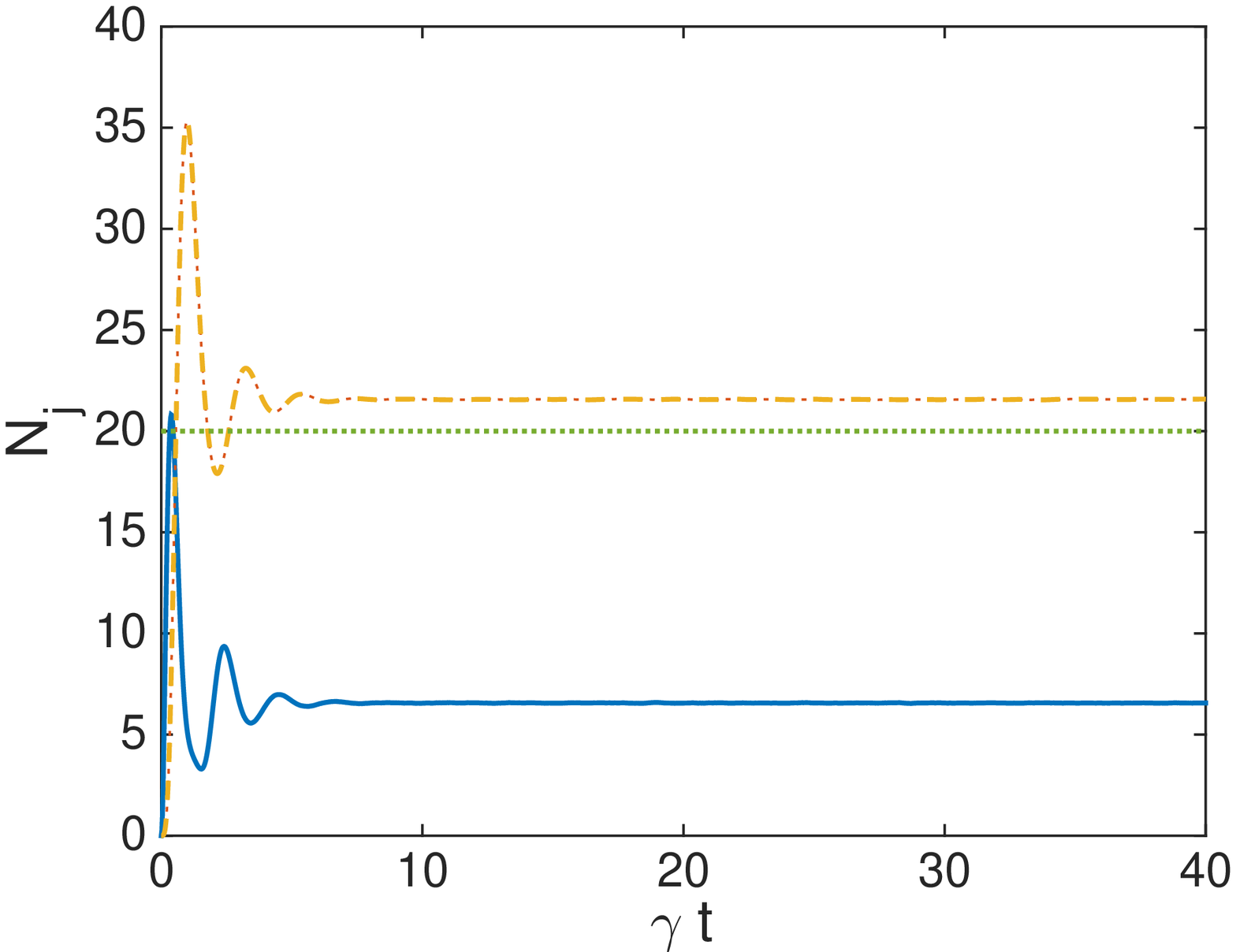}
 \end{minipage}
 \caption{(a) (Color online) The populations in each well of the triangular configuration as a function of time for $\chi=10^{-3}$ and damping at well 1. The solid line is $N_{1}$, the dash-dotted line is $N_{2}$, and the dashed line is $N_{3}$. The dotted line represents the non-interacting steady-state analytical solutions, all of which are equal for this configuration.
  \newline
 (b) The populations in each well of the triangular configuration as a function of time for $\chi=10^{-2}$ and damping at well 1. The line styles are the same as in (a).}
 \label{fig:popsg1tri}
\end{figure}

In this case the classical non-interacting steady-state solutions are found as
\begin{eqnarray}
\alpha_{1} &=& \frac{-i\epsilon}{2J-i\gamma}, \nonumber \\
\alpha_{2} &=& \frac{i\epsilon}{2J-i\gamma}, \nonumber \\
\alpha_{3} &=& \frac{i\epsilon}{2J-i\gamma},
\label{eq:classtrig1}
\end{eqnarray}
and they agree with numerical results. The numbers in all wells are the same in this case, with a $\pi/2$ phase difference between the pumped well and the other two. 
Adding in collisional nonlinearities leaves $N_{2}$ equal to $N_{3}$, while changing the actual value. For $\chi=10^{-3}$, this change is almost undetectable, as seen in Fig.~\ref{fig:popsg1tri}(a), and for $\chi=10^{-2}$, the average populations are above the non-interacting value. The interaction causes the population in well one to decrease. 

\begin{figure}[tbhp]
\centering
\begin{minipage}{.5\textwidth}
  \centering
  \includegraphics[width=.95\linewidth,height=0.65\linewidth]{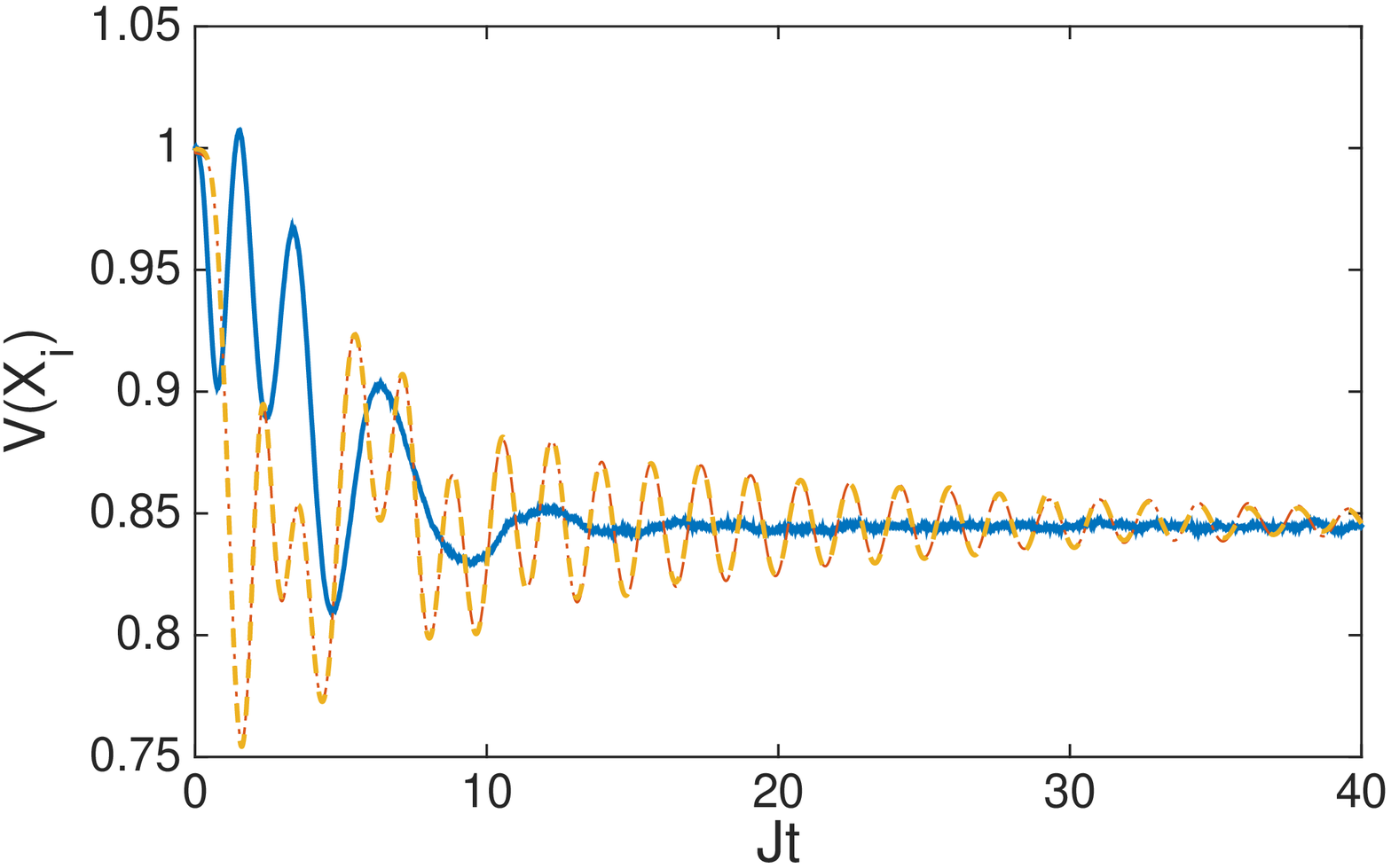}
\end{minipage}%
\begin{minipage}{.5\textwidth}
  \centering
  \includegraphics[width=.95\linewidth,height=0.65\linewidth]{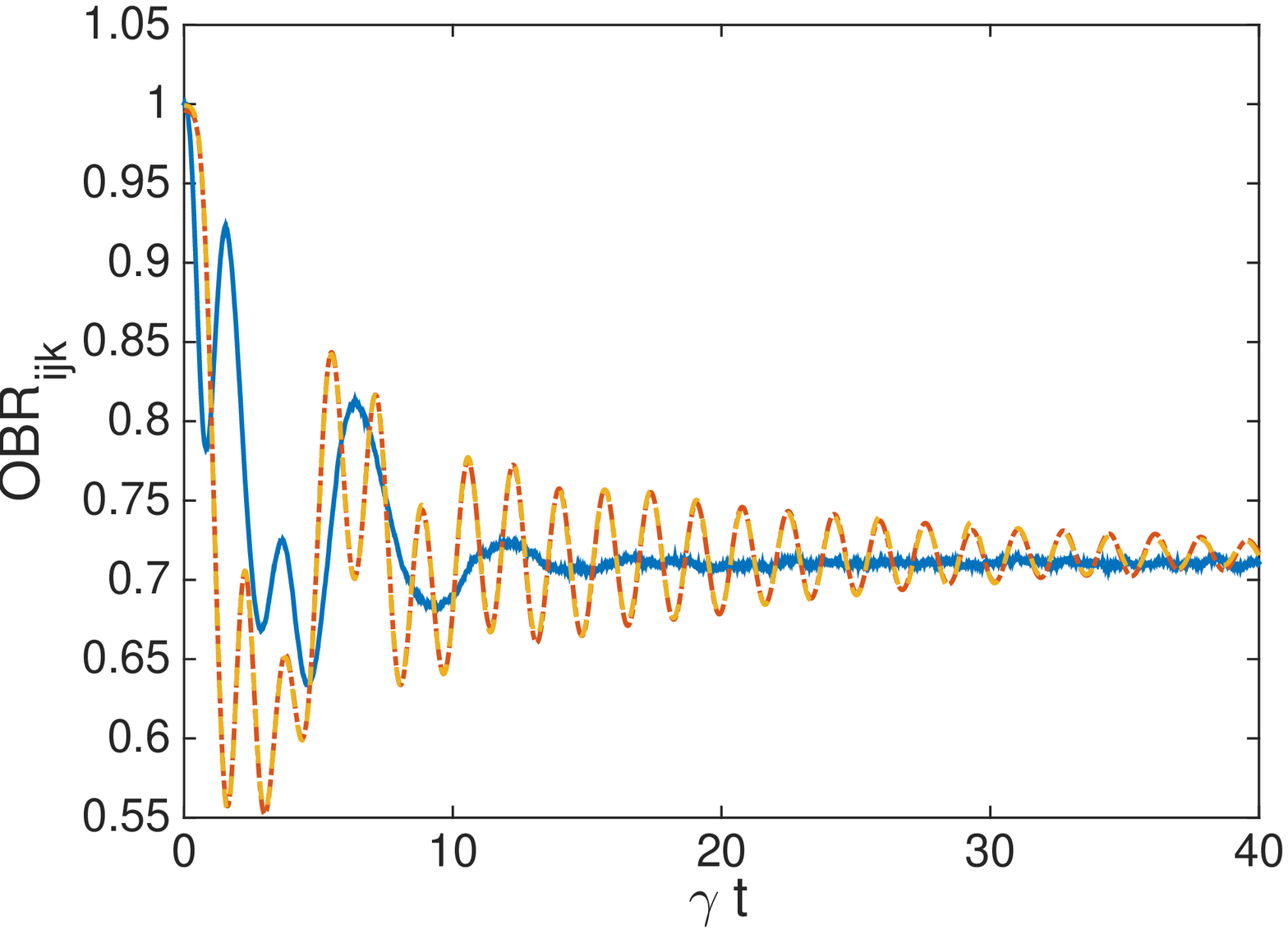}
 \end{minipage}
 \caption{(a) (Color online) The quadrature variances for pumping and damping at the same well, with $\chi=10^{-2}$. The solid line is $V(\hat{X}_{1})$ and the two indistinguishable lines are $V(\hat{X}_{2})$ and $V(\hat{X}_{3})$. All three are plotted at their optimal quadrature angle.
  \newline
 (b) The OBR correlations for $\chi=10^{-2}$ and damping at well 1. The solid line is $OBR_{123}$ while the other two are $OBR_{231}$ and $OBR_{312}$. All three are plotted at their optimal quadrature angle.}
 \label{fig:g1tricorrs}
\end{figure}

The existence of stationary states of atom number for this configuration allows us to calculate the quantum correlations of interest for both values of the nonlinearity. However, when we do the calculations we find that the correlations are time dependent for most of the criteria. The values presented in the following tables are at a fixed time, of $\gamma t = 60$ for $\chi=10^{-3}$, and $\gamma t=40$ for $\chi=10^{-2}$. We found three of the bipartite criteria, $V(\hat{X}_{1})$, $EPR_{12}$, and $EPR_{13}$, had values that did not vary with time once the intensities had reached their steady states. These are marked with asterisks in the bipartite tables below. Examples of this behaviour in the time domain are shown in Fig.~\ref{fig:g1tricorrs}, for both values of the collisional nonlinearity.

\begin{center}
\begin{tabular}{ | c || c | c | c | c | c | c |}
\hline
 \multicolumn{7}{| c |}{Bipartite entanglement, loss at 1} \\
 \hline
   &  $V(\hat{X}_{1})^{\ast}$ & $V(\hat{X}_{2})$ & $V(\hat{X}_{3})$ & $DS_{12}$ & $DS_{13}$ & $DS_{23}$  \\ 
 \hline
 \hline
$\chi = 10^{-3}$ & 0.98@124$^o$ & 0.98@142$^o$  & 0.98@142$^o$ & 3.93@135$^o$ & 3.93@135$^o$ & 3.92@142$^o$\ \\
\hline
 $\chi = 10^{-2}$ & 0.84@56$^o$  & 0.85@142$^o$   & 0.85@142$^o$   & 4.60@75$^o$  & 4.60@75$^o$  & 3.39@142$^o$ \\
 \hline
 \hline
\end{tabular}
\end{center}

We find that this system offers a small degree of quadrature squeezing and bipartite entanglement. According to the Reid criteria, the bipartite entanglement increases as the nonlinearity is increased. Once again, the Duan-Simon criteria miss entanglement that is detected by the Reid criteria. With respect to these quantum statistical criteria, this configuration produces more entanglement than that with damping at the second site. All EPR-steering found is symmetric, with either well of the chosen bipartition capable of steering the other.

\begin{center}
\begin{tabular}{ | c || c | c | c | c | c | c |}
\hline
 \multicolumn{7}{| c |}{Bipartite EPR-steering, loss at 1} \\
 \hline
   &  $EPR_{12}^{\ast}$ & $EPR_{21}$ & $EPR_{23}$ & $EPR_{32}$ & $EPR_{13}^{\ast}$ & $EPR_{31}$  \\ 
 \hline
 \hline
$\chi = 10^{-3}$ &  0.97@124$^o$  & 0.96@142$^o$   & 0.96@140$^o$   & 0.96@140$^o$  & 0.97@124$^o$  & 0.96@140$^o$ \\
\hline
 $\chi = 10^{-2}$ & 0.71@55$^o$   & 0.69@136$^o$   & 0.67@147$^o$  & 0.67@147$^o$ & 0.71@56$^o$  & 0.69@136$^o$ \\
 \hline
 \hline
\end{tabular}
\end{center}

When we examine the tripartite correlations, we find that only $V_{123}$ reaches a stationary steady state for both values of the collisional nonlinearity, with $OBR_{123}$ reaching a stationary value for $\chi=10^{-2}$. All the other correlations show persistent oscillations. The results for the vLF criteria find genuine tripartite inseparability for $\chi=10^{-3}$, but not for $10^{-2}$. The OBR criteria show that tripartite EPR-steering is possible, with all possible pairs being able to steer the results from the remaining well. The degree of violation of the inequalities increases with collisional nonlinearity.  Along with the results from the first configuration, and the results for the Reid criteria, this shows that EPR-steering inequalities are a more powerful method of detecting entanglement in this system than either the Duan-Simon or van Loock-Furusawa inequalities.

\begin{center}
\begin{tabular}{ | c || c | c | c | c | c | c |}
\hline
 \multicolumn{7}{| c |}{Tripartite entanglement, loss at 1} \\
 \hline
   &  $V_{12}$ & $V_{13}$ & $V_{23}$ & $V_{123}$ & $V_{231}$ & $V_{312}$   \\ 
 \hline
 \hline
$\chi = 10^{-3}$ & 3.94@136$^o$  & 3.94@135$^o$ & 3.93@147$^o$ & 3.96@153$^o$  & 3.97@155$^o$   & 3.97@151$^o$  \\
\hline
 $\chi = 10^{-2}$ & 4.28@87$^o$  & 4.28@87$^o$   & 3.32@136$^o$  & 4.50@160$^o$  & 4.19@167$^o$ & 4.19@167$^o$ \\
 \hline
 \hline
\end{tabular}
\end{center}

\begin{center}
\begin{tabular}{ | c || c | c | c |}
\hline
 \multicolumn{4}{| c |}{Tripartite EPR-steering, loss at 1} \\
 \hline
   &   $OBR_{123}$ & $OBR_{231}$ & $OBR_{312}$  \\ 
 \hline
 \hline
$\chi = 10^{-3}$ & 0.97@124$^o$ & 0.96@146$^o$  & 0.97@146$^o$    \\
\hline
 $\chi = 10^{-2}$ & 0.71@53$^o$  & 0.72@142$^o$ & 0.72@142$^o$\\
 \hline
 \hline
\end{tabular}
\end{center}

\section{Conclusions}
\label{sec:conclusions}

Using the truncated Wigner approximation, we have analysed the quantum statistical properties of a pumped and damped Bose-Hubbard trimer in the triangular configuration. We have found that the mean fields and the quantum correlations both depend qualitatively on the size of the collisional nonlinearity and the choice of damped well. The configuration with pumping and damping at different wells does not enter a stationary steady state of the mean fields for the lower value of nonlinearity.  For the higher value, the initial oscillations eventually damp out and it does produce both bipartite and tripartite entanglement. As is reasonably common with mixed states, the detection of these properties depends crucially on the measures chosen, with the EPR-steering criteria finding entanglement that the others missed.

The second configuration, with loss and damping at the same well, reaches a stationary steady state of the populations, even though the values of the entanglement measures generally continue to oscillate. In this case, both the Duan-Simon and Reid criteria detect inseparability and entanglement for the lower nonlinearity. For the higher nonlinearity, only the Reid criteria detect entanglement for all possible bipartitions. The tripartite OBR criteria also detect entanglement more efficiently than the vLF criteria, particularly for the higher nonlinearity.

To conclude, experimental advances in the trapping and manipulation of ultracold bosons have opened a field that is just not possible with quantum optics. Although nonlinear optical cavities can be coupled and not all cavities in an ensemble need be pumped, we cannot have optical cavities without damping. With atom traps this is possible, at least over reasonable time scales, and opens up a whole new realm of possibilities.

\section*{Acknowledgments}

This research was supported by the Australian Research Council under the Future Fellowships Program (Grant ID: FT100100515).

\end{document}